# Security analysis of an audio data encryption scheme based on key chaining and DNA encoding

Imad El Hanouti[1] · Hakim El Fadili[1]

**Abstract**

Fairly recently, a new encryption scheme for audio data encryption has been proposed in [Naskar, P.K., et al. Multimed Tools Appl (2019) 78: 25019. https://doi.org/10.1007/s11042-019-7696-z]. The cryptosystem is based on substitution-permutation encryption structure using DNA encoding at the substitution stage, in which the key generation is based on a key chaining algorithm that generates new key block for every plain block using a logistic chaotic map. After some several statistical tests done by the authors of the scheme, they claimed that their cryptosystem is robust and can resist conventional cryptanalysis attacks. Negatively, in this paper we show the opposite: the scheme is extremely weak against chosen ciphertext and plaintext attacks thus only two chosen plaintexts of 32 byte size are sufficient to recover the equivalent key used for encryption. The cryptosystem's shuffling process design is vulnerable which allow us recovering the unknown original plaintext by applying repeated encryptions. Our study proves that the scheme is extremely weak and should not be used for any information security or cryptographic concern. Lessons learned from this cryptanalytic paper are then outlined in order to be considered in further designs and proposals.

## 1. Introduction

Due to some intrinsic characteristics of multimedia data (videos, audios…) [1], including bulk data size, strong correlation between uncompressed neighbor data-units, high presence in real-time systems and the necessity for fast processing and transmission [2-3]. Multimedia data processing and transmission has become a challenging area of research in order to convoy modern applications requirements [4]. Information security and data privacy constitutes a crucial requirement in modern network applications and data transmission. Public networks are threatened by several cyber-attacks due to the high and random accessibility to computer networks. Multimedia data security as any data type security is one of the most important process to handle before establishing any information transfer over public networks. The necessity for fast multimedia processing applications that responds to real-time requirements makes multimedia application development a hot topic in recent research arena [5-9].

---

[1] I. El Hanouti (✉) · H. El Fadili
Computer Science and Interdisciplinary Physics Laboratory (LIPI), National School of Applied Sciences, SMBA University, Fez, Morocco
Email : imad.elhanouti@usmba.ac.ma

Cryptography and cryptographic protocols [10] make one basic layer for any data type security, cryptography is the art of designing highly secure systems that aims to convert original data from a readable state to an unrecognizable form except for entities that own the secret key. Conventional cryptosystems has been shown in some related works to be insufficient to respond to all mentioned requirements [11]. That's why designing new cryptosystems with high security and fast processing time has become a primordial issue since the beginning of our current century [12-15]. Two of the most relevant requirements in modern cryptography are *confusion* and *diffusion* depicted by C. Shannon in his paper "Communication Theory of Secrecy Systems" in 1949 [16]. Confusion refers to making the relationship between the key and the ciphertext as complex as possible while diffusion refers to the property that the redundancy in the statistics of the plaintext is *dissipated* in the statistics of the ciphertext. It has been observed that chaos theory exhibits similar proprieties as confusion and diffusion in cryptology theory [17].

Chaos theory is a branch of mathematics studying the strange behaviors of dynamic systems and chaotic attractors that has two main characteristics [17]: the sensitivity to initial conditions (small changes in initial conditions leads to significantly great changes in the orbit of the strange attractor), and ergodicity. Chaos theory has been widely used in designing new cryptosystems [5-8,13-15]. Nevertheless, the security level claimed by authors of each proposal is still in doubt until carrying out a sound cryptanalytic study on their design [20]. Some of them have been already cryptanalyzed in subsequent works [18-19]. As a result, many recommendations and guidelines have been proposed to assess some basic level of security in new works [17,21].

From multimedia to big data, the need for huge and complex calculation capacity and storage space pushes researchers to look up for other computing capabilities [22] and environments that meet all requirements for computing-world tendency and modern applications. DNA computing is a rising filed of new computing generation which makes use of biochemical reactions techniques to perform computations across a medium of biological DNA molecules [23]. DNA computing has three main advantages over traditional silicon computing, which are inherited basically from some intrinsic characters of DNA molecules: a high degree of parallelism, low energy consumption and a vast amount of information storage [24].

In [25] a new proposed cryptosystem dedicated especially for audio data security is designed based on DNA encoding and the logistic chaotic map. The new proposed system is a block cipher that operates on blocks of 32 byte size. It is based on a key chaining algorithm that generates different new key block for any plain block. The scheme's design is based on two main operations: substitution using a DNA XOR (exclusive-or) operation, followed by a shuffling process to permute bytes within each block. The experimental and statistical tests done by the authors of the scheme shows good performances [25]. However, in this paper we scrutinized the new proposed system and we found that the scheme has some critical flaws. Exploiting those vulnerabilities allows us to conduct many cryptanalytic attacks on the overall system as we will demonstrate in further sections.

The organization of this paper is as follow: the second section describes the audio data cryptosystem under study. Section 3 outlines some vulnerabilities and security flaws found in the scheme, two main vulnerabilities are considered as critical flaws and make the system weak under many cryptanalytic attacks. In section 4 we illustrate our proposed chosen plaintext attack as well as a small discussion about chosen ciphertext attack. In section 5 we illustrate our so called *cycle attack* which is a customized version of chosen plaintext attack that aims to recover the plaintext without doing any programming effort and only by exploiting the cryptosystem's encryption machinery. In section 6 we depict some lessons learned from this cryptanalytical paper to avoid designing systems with same vulnerabilities in further proposals. Finally, the section 7 discusses our results and concludes the paper.

## 2. Description of the audio data encryption scheme

The scheme under study [25] is labeled in this paper as DECS-AU scheme (DNA Encoding and Chanel Shuffling Audio Encryption scheme). It is based on the logistic chaotic map [26] and DNA encoding to handle different operations of the process of encryption as well as decryption. The DECS-AU is a block cipher that operates on blocks of 32 byte length. In this section we re-describe briefly all the process of the encryption operation as detailed in [25] followed by a tiny description of the decryption process. DECS-AU could be described in three basic steps, which are: *key scheduling and chaining*, *substitution operation* and *shuffling operation*.

1) *Key scheduling and chaining*:
   Each single plain block[2] $b_i$ of 32 byte size is encrypted with a corresponding key $K_i$ derived basically from the previous key block $K_{i-1}$ using the logistic chaotic map Eq. (2). The *original* 32 byte key (the one chosen by the user) is denoted as $MK$ (master key) and used to generate the first key block $K_1$ for the encryption of the first plain block $b_1$. As mentioned in [25], consecutive blocks use definitely distinct keys. The key $K_i$ is generated using the previous generated key $K_{i-1}$ as follow:

$$K_i(j) = mod\left(\lfloor x_j \times \left(K_{i-1}^2(j) + K_{i-1}(j-1)\right)\rfloor, 256\right), j = 1,2,..,32 \quad (1)$$

Where $K_i(j)$ is the $j-th$ byte of the key $K_i$, $mod(x, 256)$ function denotes reduction modulo 256, the floor function $\lfloor . \rfloor$ calculates the greatest integer less than or equal $(.)$, we define $K_{i-1}(0)$ as: $K_{i-1}(0) = K_{i-1}(1)$, and we define $K_0$ as $K_0 = MK$, finally $x_j$ is defined using the logistic map equation as :

$$x_j = 3.99 \times x_{j-1} \times (1 - x_{j-1}) \quad (2)$$

Where $x_0$ is calculated by the following equation:

$$x_0 = \frac{\sum_{j=1}^{32} MK(j) \times (j+1)}{2^{21}} \quad (3)$$

That is, the initial condition $x_0$ is calculated first from the *master key MK*, and then the logistic map with a fixed parameter $\lambda = 3.99$ is iterated according to the position of the key byte.

The last operation in key scheduling process is to update the generated values according to the following updating operation:

$$K_i(j) = \begin{cases} K_i(j), & if \ K_i(j) \neq K_{i-1}(j) \\ \lfloor x_j \times 256 \rfloor, & if \ K_i(j) = K_{i-1}(j) \end{cases} \quad (4)$$

This updating operation ensures that two successive key blocks are byte-wise distinct one from another. This *key chaining* ensures a good diffusion of the master key $MK$ along all the key-blocks.

---

[2] Some notations from the original paper are modified without affecting its main meaning.

2) *Substitution operation*:

Substitution operation aims to confuse the plaintext[3] using the key generated at the key scheduling stage. Here, the same process for one block is done for every block $b_i$ of the plaintext $P$. The substitution operation could be described in three steps:

1) DNA encoding operation:

Each byte from the block $b_i$ denoted as $b_i(j)$ (the $j-th$ byte of the plain block $b_i$) is encoded into its DNA sequence according to a well-defined rule $r_j$ calculated as:

$$r_j = mod\left(\left|x_j \times \left(K_i^2(j) + K_i(j-1)\right)\right|, 8\right) \quad (5)$$

The encoding map used in [25] is depicted in Table 1. The same process is done to encode the entire key $K_i$ to its corresponding DNA nucleotide sequence. The same rule is used to encode the same $j-th$ byte from both $b_i$ and $K_i$. We denote the encoding function that encodes a single byte $d$ according to the rule $r_j$ as:

$$DNAe_{r_j}(d) = N_4 N_3 N_2 N_1, \ N_i \in \{'A','T','G','C'\} \quad (6)$$

**Table 1** DNA encoding and decoding map according to the rule type used in [25]

| Rule $r_j$ | A | T | G | C |
|---|---|---|---|---|
| $r_j = 1$ | 00 | 11 | 10 | 01 |
| $r_j = 2$ | 00 | 11 | 01 | 10 |
| $r_j = 3$ | 11 | 00 | 10 | 01 |
| $r_j = 4$ | 11 | 00 | 01 | 10 |
| $r_j = 5$ | 10 | 01 | 00 | 11 |
| $r_j = 6$ | 10 | 01 | 11 | 00 |
| $r_j = 7$ | 01 | 10 | 00 | 11 |
| $r_j = 8$ | 01 | 10 | 11 | 00 |

2) DNA XOR operation:

After the encoding of bytes using DNA encoding rules, DNA XOR operation is carried out between block's DNA codes and key's DNA codes. In [25] the definition of DNA XOR operation is based on the binary XOR operation (bitwise EX-OR operation): for any given rule the DNA code for a DNA XOR operation is defined as the DNA encoded bitwise XOR operation of the two operands in binary form. Table 2 depicted the DNA XOR operation according to the rule $r_j = 1$.

**Table 2** DNA XOR operation according to rule 1 [25]

| DNA XOR ⊞ | A | T | G | C |
|---|---|---|---|---|
| A | A | T | G | C |
| T | T | A | C | G |
| G | G | C | A | T |
| C | C | G | T | A |

3) DNA decoding operation:

After the calculation of the DNA XOR operation between key entities and block entities, the results are decoded back to the binary form according to the same rule used for encoding of each byte. The decoding function according to the rule $r_j$ decodes a quadruplet of nucleotides back to a single binary byte $d$ and it is denoted as:

---

[3] The two terms "plaintext" and "plain blocks" are used alternatively in this paper, the same for "keys" with "key blocks" and "ciphertext" with "cipher blocks"

$$DNAd_{rj}(N_4N_3N_2N_1) = d, \ N_i \in \{'A','T','G','C'\} \tag{7}$$

The overall substitution process could be formed mathematically as:

$$S_i(j) = DNAd_{rj}\left(DNAe_{rj}(b_i(j)) \boxplus DNAe_{rj}(K_i(j))\right), \ j = 1,\ldots,32 \tag{8}$$

Where $S_i(j)$ is the $j-th$ byte of the $i-th$ substituted block, $DNAd_{rj}$ and $DNAe_{rj}$ are previously defined in Eqs. (6) and (7) and $\boxplus$ denotes DNA XOR operation.

3) *Shuffling operation*:
   The last operation in the encryption process is the shuffling of bytes within each substituted block $S_i$. The shuffling algorithm calculates shuffling indexes, which is a permutation map for each substituted block and it depends on the corresponding key block. Odd indexes are stored in one array called right channel $Rc = \{Rc(j) \in 2\mathbb{N} + 1\}_{j=1}^{16}$ and even indexes are stored in another array called left channel $Lc = \{Lc(j) \in 2\mathbb{N}\}_{j=1}^{16}$. And every byte $S_i(Lc(j))$ is swapped with its corresponding byte $S_i(Rc(j))$. The algorithm used to generate the two left and right channels is depicted in Fig. 1. The shuffling operation is mathematically equivalent to:

$$C_i(j) = S_i(j^*) \tag{9}$$

Where $C_i(j)$ is the $j-th$ byte of the $i-th$ cipher block, and $j^*$ is defined as

$$j^* = f(j) \tag{10}$$

Where $f$ is a bijective function from the set $P$ to $P$ such that $P = Rc\frown Lc = \{p_i\}_{i=0}^{31}$ where $A\frown B$ denotes the concatenation of the the set $A$ and the set $B$. $f$ is defined as:

$$f(p_i) = p_{mod(i+16,32)}, \ p_i \in P \tag{11}$$

According to Kirchhoff's principle [27], the underlying algorithm and all design of a cryptographic system should be known to public and the secrecy of the system must rely only on the secret key. The overall encryption process could be formulated mathematically as:

$$C_i(j) = DNAd_{rj*}\left(DNAe_{rj*}(b_i(j^*)) \boxplus DNAe_{rj*}(K_i(j^*))\right), \ j = 1,\ldots,32 \tag{12}$$

Where $C_i(j)$ is the $j-th$ byte of the $i-th$ cipher block, and $j^*$ is defined in Eq. (10).

The decryption process is done by taking encryption steps in reverse order. Firstly, using the same key chaining process to de-shuffle the cipher block, and then de-mask it according to the same substitution operation.

```
1:   INPUT the current 32 element key block array K
2:   OUTPUT two arrays of 16 elements Rc and Lc
3:   START
4:   READ K
5:   CREATE array V having size = 32
6:   CREATE array Rc having size = 16
7:   CREATE array Lc having size = 16
8:   SET p1 to 0
9:   SET p2 to 0
10:  FOR each element in K
11:    REPEAT
12:      SET t to (element Mod 32)
13:      INCREMENT element
14:    UNTIL v[t]=0
15:    IF (t Mod 2)=0 THEN
16:      SET Rc[p1] to (t+1)
17:      INCREMENT p1
18:    ELSE
19:      SET Lc[p2] to (t+1)
20:      INCREMENT p2
21:    ENDIF
22:    SET v[t] to 1
23:  ENDFOR
24:  STOP
```

**Fig. 1** Shuffling process indexes generation algorithm

## 3. Security flaws in DECS-AU scheme

In this section we outline some critical observed vulnerabilities in the DECS-AU scheme. Some vulnerabilities affect directly the overall system's security, while others are considered as weaknesses of design and *bad* implementation of some ideas and operations, therefore, they could be exploited to enhance some attacks or partially reveal some secret parts of data or keys. Our observed security flaws are described as follow:

1) Bad confusion/diffusion [16] resulting from the fact that in key chaining operation only the previous key is considered, neither the plain blocks nor the cipher blocks are related to the key generation and chaining process. This is a crucial vulnerability in the system, knowing only the first key block will immediately allow us to recover any desired key block by application of the known algorithm of the key scheduling and chaining operation as described in the previous section. This vulnerability opens the door to many cryptanalytic attacks (e.g. chosen plaintext attack (Section 4.1), chosen ciphertext attack (Section 4.2), differential attack (section 4.1), known plaintext attack …).
   ***Proposition 1:*** *Since the key blocks in DECS-AU are related only to the previous key block, then knowing the first key block $K_1$ is sufficient to recovering all the keys $K_i$ used for the encryption of any plaintext using the same master key $MK$.*
   According to this proposition, we will focus in our attacks to recover only the first key block.
2) No security effect is added by DNA encoding: DNA computing is a rising field of computing where the medium of operations is biological DNA molecules which adds more benefits to the world of computing (due to some intrinsic molecule's characteristics) such that high degree of parallelism, low energy consumption and vast storage space [24]. In last years, several proposals has been published to implement some computing capabilities using DNA strands such that making logic gates [28,29] and implementing arithmetic operations [30,31]. However in the DECS-AU scheme, the methodology adopted to encode binary data and perform DNA XOR operation is by definition based on binary XOR operation. Thus, DNA operations are no more than just and encoding/decoding operations. The definition of the DNA XOR operation as explicitly depicted in [25] could be equivalently formulated for any given rule $r_j$ as:

$$DNAd_{r_j}\left(DNAe_{r_j}(d_1) \boxplus DNAe_{r_j}(d_2)\right) = d_1 \oplus d_2 \tag{13}$$

Where $d_1$ and $d_2$ are two bytes.

From this last equation we can say that the DNA XOR operation as done in [25] is equivalent to a simple bitwise EX-OR operation in the binary form (denoted as $\oplus$). That is, this operation *may* benefits only from some DNA computing pros including parallelism, vast storage space ... And no security effect is envisaged to be added by this mean.

3) *Period-2* shuffling map (shuffling an element two times regains its original position): the shuffling operation is a permutation process that shuffles bytes inside each substituted block $S_i$. Generally, the total number of possible permutations for a block of 32 *distinct* bytes is $n = 32! \approx 2^{118}$ permutations, and the range of a permutation, which is the smallest integer $R$ that if the permutation is applied $R$ times the data will regain its initial form (no permutation occurred) should be as large as possible and close to $n$. The algorithm (shown in Fig. 1) used by authors of the DECS-AU to generate permutation maps is very weak and all permutations generated by the algorithm are of range $R = 2$. This critical vulnerability makes all the permutation maps (that could be considered as permutation keys) a weak keys, and makes the overall system weak against cycle attack (to be detailed more in Section 5).

***Proposition 2:*** *the range R of all permutation maps (shuffling) in [25] is $R = 2$. That is, for any $p_n \in P, f(f(p_n)) = p_n$. In other words, for any block $b_i$, the application of only shuffling process twice has no effect on the block $b_i$.*

***Proof:***

Let two elements $p_n, p_m \in P$, such that $f(p_n) = p_m$, we have to prove that $f(f(p_n)) = p_n$:

From Eq. (11) we have:

$$f(f(p_n)) = f(p_m) = p_{mod(m+16,32)} \tag{14}$$

And:

$$m = mod(n + 16,32) \tag{15}$$

Thus:

$$\begin{aligned} mod(m + 16,32) &= mod(mod(n + 16,32) + 16,32) \\ &= mod(n + 16 + 16,32) \\ &= mod(n + 32,32) \\ &= n, \quad \text{since } n \in \mathbb{N}/32\mathbb{N} \end{aligned}$$

Thus, we have proved that for any given $p_n \in P$:

$$f(f(p_n)) = p_n \tag{16}$$

This proves the proposition.

∎

4) Fixed $\lambda$ parameter of the chaotic logistic chaotic map: the parameter $\lambda$ of the chaotic logistic map is by definition in the range [0,4] of real numbers. However, the logistic map exhibits chaotic behavior only when the parameter $\lambda$ belongs to some distinct regions within [0,4] but not all the range. The determination of these *chaotic regions* needs deep analyzing of the bifurcation diagram to exclude all non-chaotic *periodic windows*. Nevertheless, in [25] only one fixed parameter $\lambda = 3.99$ is used in every call for the chaotic map, which may allow one to deduce the value of the initial condition $x_0$ from any value $x_k$ of the chaotic map and the order of iterations $k$.

5) No padding method is defined for the DECS-AU scheme to adjust the last block to a 32 byte block: if the padding of blocks is based on *zero padding* (use null bytes to pad), then information about the last key block will be clearly manifested (in shuffled form) in the last cipher block.
6) The applicability of the DNA XOR operation as defined in [25] is seriously questionable (The XOR DNA is dependent on the rule at each operation). DNA XOR gates such as in [29] are based on oligonucleotides strands, thus every binary state (0 or 1) is encoded using an oligonucleotide strand, which allows making some DNA logic gates. The feasibility of the DNA XOR operation as defined in [25] should be at least simulated.

## 4. Differential chosen plaintext attack and chosen ciphertext attack

### 4.1 Differential chosen plaintext attack

Chosen plaintext attack is an attack in which the attacker gains temporary access to the encryption machinery. Thus, he could choose some intentionally made plaintexts and get their corresponding ciphertext under some unknown key. The objective of the attack is to recover the key or the equivalent key used for encryption. From (Proposition 1), if we recovered the first key block $K_1$, all subsequent keys could be recovered. In this attack, we will demonstrate that only two chosen plaintexts of 32 byte size are sufficient to recover the first key block $K_1$ and thus recovering the full equivalent key used in the encryption process.

In this section we denote the first plaintext as $b_1$ and the second plaintext as $b_2$ both are blocks of 32 bytes. Their corresponding substituted blocks are denoted as $S_1$ and $S_2$ respectively. And their corresponding ciphertexts are denoted as $C_1$ and $C_2$ respectively. Fig. 2 illustrates a flowchart of our proposed differential chosen plaintext attack. Our proposed attack could be described in two main steps:

1) Recovering the shuffle map of the first block:
   In order to recover the shuffle map of the first block, we have to satisfy two conditions:
   i) Make unique identifier for each byte position in the plaintext block
   ii) Cancel out the substitution process, thus $S_i = b_i, i = 1,2$.

   By this mean, we will observe the new position of each byte in the ciphertext which will allow us to deduce the shuffling map. In order to achieve that, we will combine a differential attack with a chosen plaintext one to cancel out the substitution effect.

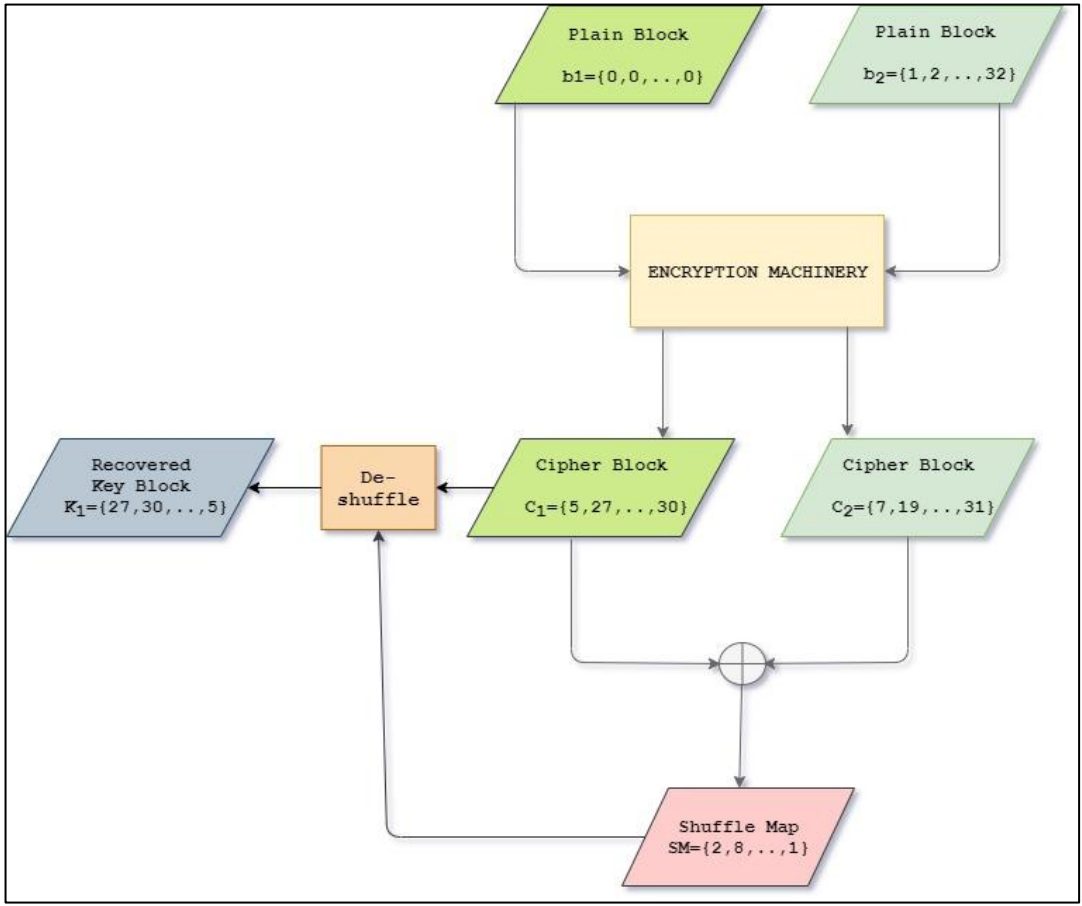

**Fig. 2** General schema of our proposed differential chosen plaintext attack

The substitution process is based on the DNA XOR operations, and from Eq.(13) we have that DNA XOR operation is equivalent to binary bit-wise XOR operation. Let us consider the difference between the two ciphertext blocks as:

$$\Delta(C_1, C_2) = C_1 \oplus C_2 \qquad (17)$$

From Eq. (12) we have for $j = 1,\ldots,32$:

$$\Delta(C_1,C_2)(j) = DNAd_{rj*}\left(DNAe_{rj*}(b_1(j^*)) \boxplus DNAe_{rj*}(K_1(j^*))\right) \oplus DNAd_{rj*}\left(DNAe_{rj*}(b_2(j^*)) \boxplus DNAe_{rj*}(K_1(j^*))\right) \qquad (18)$$

Where $j^*$ is defined in Eq. (10).

Thus, using Eq. (13) we have:

$$\Delta(C_1,C_2)(j) = b_1(j^*) \oplus K_1(j^*) \oplus b_2(j^*) \oplus K_1(j^*) = b_1(j^*) \oplus b_2(j^*) \qquad (19)$$

Thus, we could write the difference in ciphertext blocks in function of the difference in plaintext blocks as:

$$\Delta(C_1,C_2)(j) = \Delta(b_1,b_2)(j^*) \qquad (20)$$

Where $j^*$ is defined in Eq. (10).

In other words, we do cancel out the substitution effect and only shuffling effect remains. By this mean, we have satisfied the second condition mentioned in the beginning of this sub-section and we could easily deduce the permutation map if we choose carefully the two plaintexts $b_1$ and $b_2$ to satisfy the first condition in such a way to have unique identifier value for each byte in $\Delta(b_1, b_2)$ plain block.

One possible proposition is to make the plain blocks difference as:

$$\Delta(b_1, b_2)(j) = j, \ j = 1,2,\ldots,32 \tag{21}$$

The two chosen plaintexts could be chosen as:

$$\begin{cases} b_1(j) = 0 \\ b_2(j) = j \end{cases}, \ j = 1,2,\ldots,32 \tag{22}$$

The shuffling map for the first block could be recovered according to Eq. (10) as:

$$P = \{p_i\}_{i=0}^{31} = \{\Delta(b_1, b_2)(j)\}_{j=1}^{16} \frown \{\Delta(C_1, C_2)(j)\}_{j=1}^{16} \tag{23}$$

2) Recovering the key block $K_1$:

After the recovery of the shuffling map, the recovery of the key $K_1$ is quite simple operation. From Eqs. (12), (13) and (22) we have:

$$C_1(j) = b_1(j^*) \oplus K_1(j^*) = 0 \oplus K_1(j^*) = K_1(j^*) \tag{24}$$

Thus, $C_1$ is a shuffled version of the key block $K_1$. According to (Proposition 2), we simply apply the shuffling map (recovered from the previous step) once on the ciphertext $C_1$ to recover the key block $K_1$. Generally according to (Proposition 2) we have:

$$C_1(j) = K_1(j^*) \Leftrightarrow K_1(j) = C_1(j^*) \tag{25}$$

Where $j^*$ is defined in Eq. (10).

3) Recovering all key chains:

After the recovery of the first key block $K_1$, and according to (Proposition 1), we use the algorithm described in the key scheduling and chaining algorithm from the (Section 2) to recover as many key blocks as needed to decrypt any other plaintext encrypted under the same master key $MK$.

Fig. 3 demonstrates our proposed differential chosen plaintext attack: the time-domain plots of the two chosen plaintexts $b_1$ and $b_2$ are shown respectively in Fig. 3 (a) and (b) and their corresponding ciphertexts time-domain plots are shown respectively in Fig. 3 (c) and (d). The frequency-domain plot of the *supposed unknown* plaintext (a sine wave of frequency $F = 700Hz$) is shown in Fig. 3 (e) and its corresponding ciphertext's frequency-domain plot is shown in Fig. 3 (f). All the samples are encrypted under the same master-key:

$MK = \{D6, D0, 75, 2A, 85, 80, D9, 3B, D4, CE, 20, FA, 78, 5D, 5E, A0, CD, 1D, 17, 0D, 51, B2, 37, E7, 40, 52, AA, 6E, F1, 71, 60, CE\}$ .

Note that all keys used in this paper are generated randomly using *rand* function from *octave*[4], and that all samples are mono-channel samples that are sampled under a sampling frequency of $fs = 8000Hz$ using 8-bits per sample. The recovered key is used to decrypt the ciphertext which its frequency-domain plot is shown in Fig. 3 (f) and the time-domain (portion of it) as well as the frequency-domain plots of the recovered plaintext are shown respectively in Fig. 3 (g) and (h). The recovered plaintext matches the original one.

---

[4] Octave freeware under GNU GPL license at: https://gnu.org/software/octave/

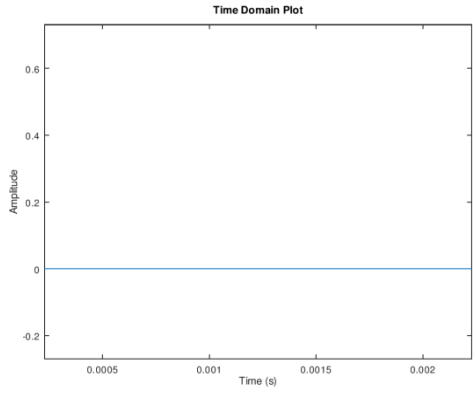

(a)

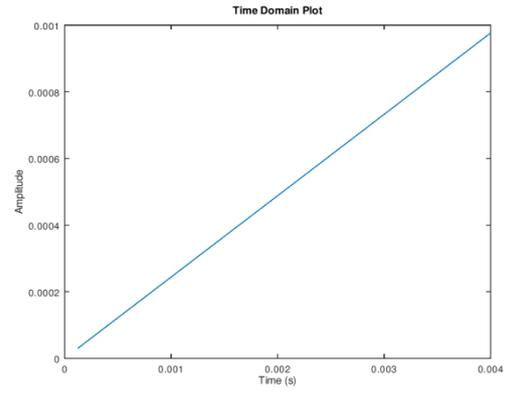

(b)

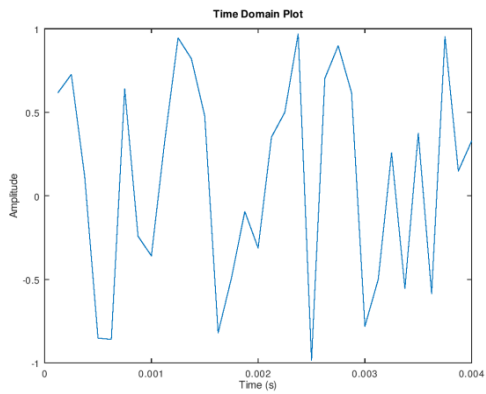

(c)

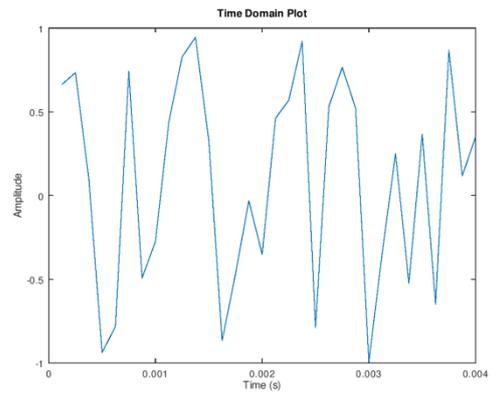

(d)

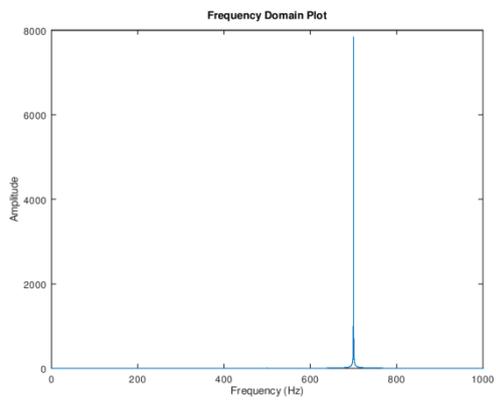

(e)

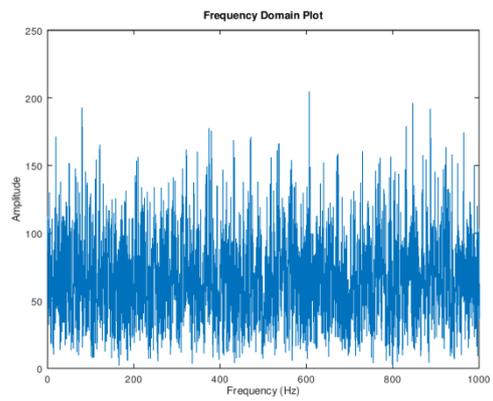

(f)

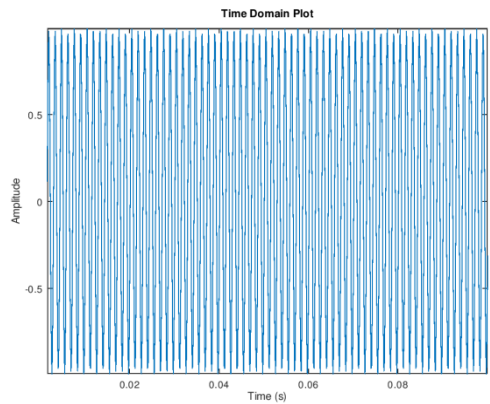

(g)

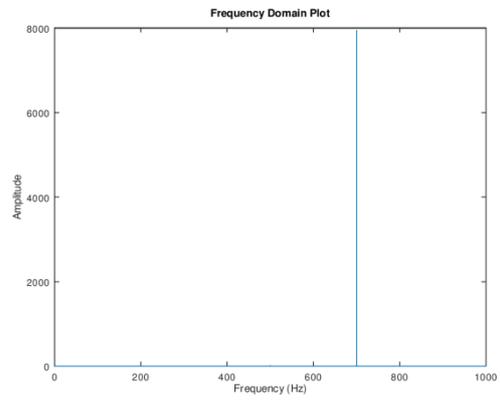

(h)

**Fig. 3** Demonstration of our proposed differential-chosen plaintext attack: (a) time-domain plot of the first chosen plaintext (b) time-domain plot of the second chosen plaintext (c) time-domain plot of the encrypted first chosen plaintext (d) time-domain plot of the encrypted second chosen plaintext (e) frequency-domain plot of the unknown targeted plaintext (f) frequency-domain plot of the encrypted unknown targeted plaintext (g) portion of time-domain plot of the recovered plaintext (h) frequency-domain plot of the recovered plaintext

### 4.2 Chosen ciphertext attack

For a chosen ciphertext attack, only one chosen ciphertext of 32 bytes size is required to directly recover the first key block $K_1$. The flowchart of our proposed chosen ciphertext attack is illustrated in Fig. 4. The chosen cipher block is $C = \{C(j) = 0\}_{j=1}^{32}$, thus shuffling process decryption will not affect the cipher block since all bytes are with the same value, thus, the substituted block is itself the cipher block: $S = C$.

Then, according to Eqs. (12) and (13) we have:

$$C(j) = b(j) \oplus K_1(j) = 0, \quad j = 1,2,\ldots,32 \tag{26}$$

Where $b$ is the corresponding plaintext of the ciphertext $C$, Thus:

$$b(j) = K_1(j), \quad j = 1,2,\ldots,32 \tag{27}$$

That is, the key block $K_1$ is manifested directly in the decrypted plaintext $b$.

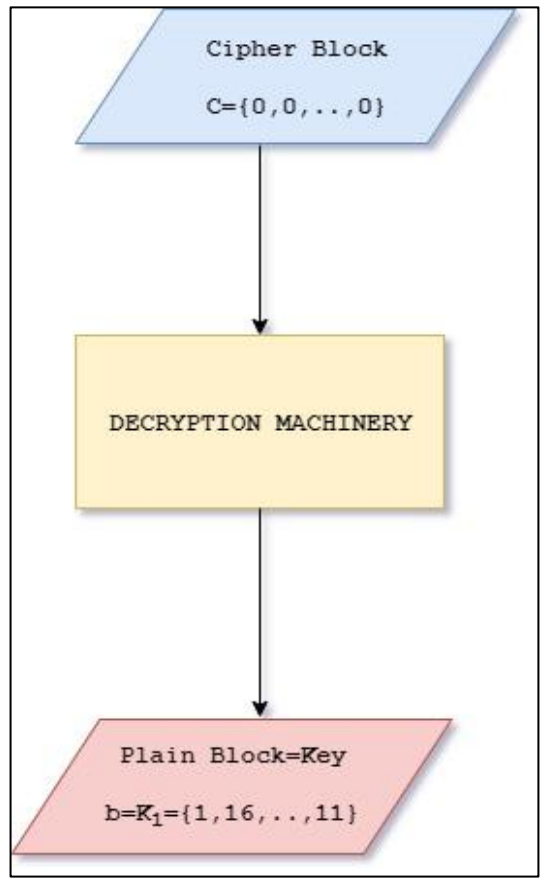

**Fig. 4** General flowchart of our proposed chosen ciphertext attack

## 5. Weak keys and cycle attack

Weak keys are keys that lead to some undesired results (e.g. no or *bad* encryption, self-decryption…). That is, if for example encrypting some plaintext with the same key twice leads to recovering the original plaintext, then the key is considered as a weak key. We extend this definition to our proposed *cycle attack* to consider a *relatively small* number of successive encryptions $c \geq 2$.

Our proposed cycle attack is based on repeating encryptions of a particular unknown plaintext $c$ times under the same unknown key $K$, Fig. 5 shows the general flowchart of our proposed cycle attack. It could be seen as a chosen plaintext attack with $c - 1$ chosen plaintext. The cycle attack could success (e.g. be feasible) only under *weak keys,* that's why for any cryptographic system it is better worth to exclude all weak keys from the key space.

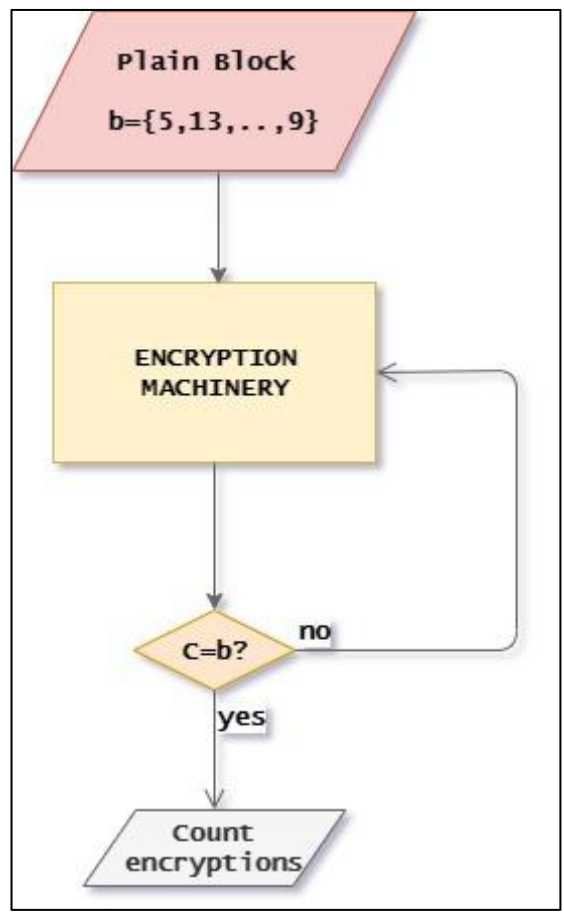

**Fig. 5** General flowchart of our proposed cycle attack

From (Proposition 2), we know that the range (cycle) of all *shuffling keys* is equal to two: $R = 2$. Thus, for any key, if the encryption process is applied twice, the shuffling permutation will be canceled.

**Lemma 1:** *for any even number $R \in 2\mathbb{N}$, applying R successive DECS-AU encryptions on the same plaintext will cancel out the shuffling process effect.*

**Proof:** trivially proved using (Proposition 2), by applying $R/2 \in \mathbb{N}$ double encryptions. ∎

From Eq. (13), the substitution process is based on XOR operation. Bit-wise XOR operation is characterized by the fact that for any byte $a$ we have the following propriety: $a \oplus a = 0$. Thus, in order to cancel out the effect of XOR operation we have to apply it twice.

***Proposition 3:*** *Under any master key K, the DECS-AU scheme is weak under our described cycle attack. And only $R = 4$ application of successive encryptions is sufficient to recover the unknown original plaintext.*

***Proof:***

Let $b$ be a plain block and $C$ is its corresponding cipher block under some key $K$, $b(k)$ is the $k - th$ byte of $b$, we have to prove that after 4 encryptions we have: $C^4(k) = b(k)$ where $C^i$ is the cipher block after $i$-encryptions.

We have according to Eqs. (12), (13) and (25):

$$C^1(k^*) = b(k) \oplus K(k) \tag{28}$$

$$C^2(k) = C^1(k^*) \oplus K(k^*) = b(k) \oplus K(k) \oplus K(k^*) \tag{29}$$

$$C^3(k^*) = C^2(k) \oplus K(k) = b(k) \oplus K(k) \oplus K(k^*) \oplus K(k) = b(k) \oplus K(k^*) \tag{30}$$

$$C^4(k) = C^3(k^*) \oplus K(k^*) = b(k) \oplus K(k^*) \oplus K(k^*) = b(k) \tag{31}$$

That is, we have proved that $C^4 = b$ wich implies that after 4 encryptions under the DECS-AU scheme we will recover the original plain.

∎

To experimentally demonstrate this proposed attack, we consider an unknown plaintext[5] where both time-domain and frequency-domain plots are shown in Fig. 6 (a) and (b) respectively. Fig. 6 (c) and (d) represent its time-domain and frequency-domain plots respectively after one encryption. Fig. 6 (e) and (f) represent time-domain and frequency-domain plots respectively after two encryptions. Fig. 6 (g) and (h) represent time-domain and frequency-domain plots respectively after three encryptions. And Fig. 6 (i) and (j) shows the time-domain and frequency-domain plots respectively of the 4-th encryption operation. As shown, the 4-th encryption returns back to the original plaintext, which demonstrates experimentally our described attack. The master key used for these encryptions is:

$$MK = \{AD, 90, 41, AE, E7, FE, 1F, 4C, 36, 3A, 02, 5C, 37, 54, 51, AD, 95, 76, 1F, D1, C6, 1E, 7B, 66, AC, DE, 33, D4, 25, FB, 52, 80\}$$

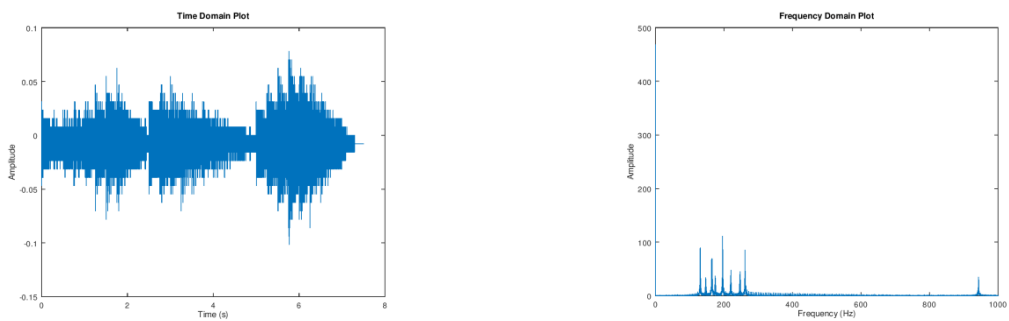

---

[5] The audio sample used for this illustration is available freely under:
https://archive.codeplex.com/?p=audiotestfiles

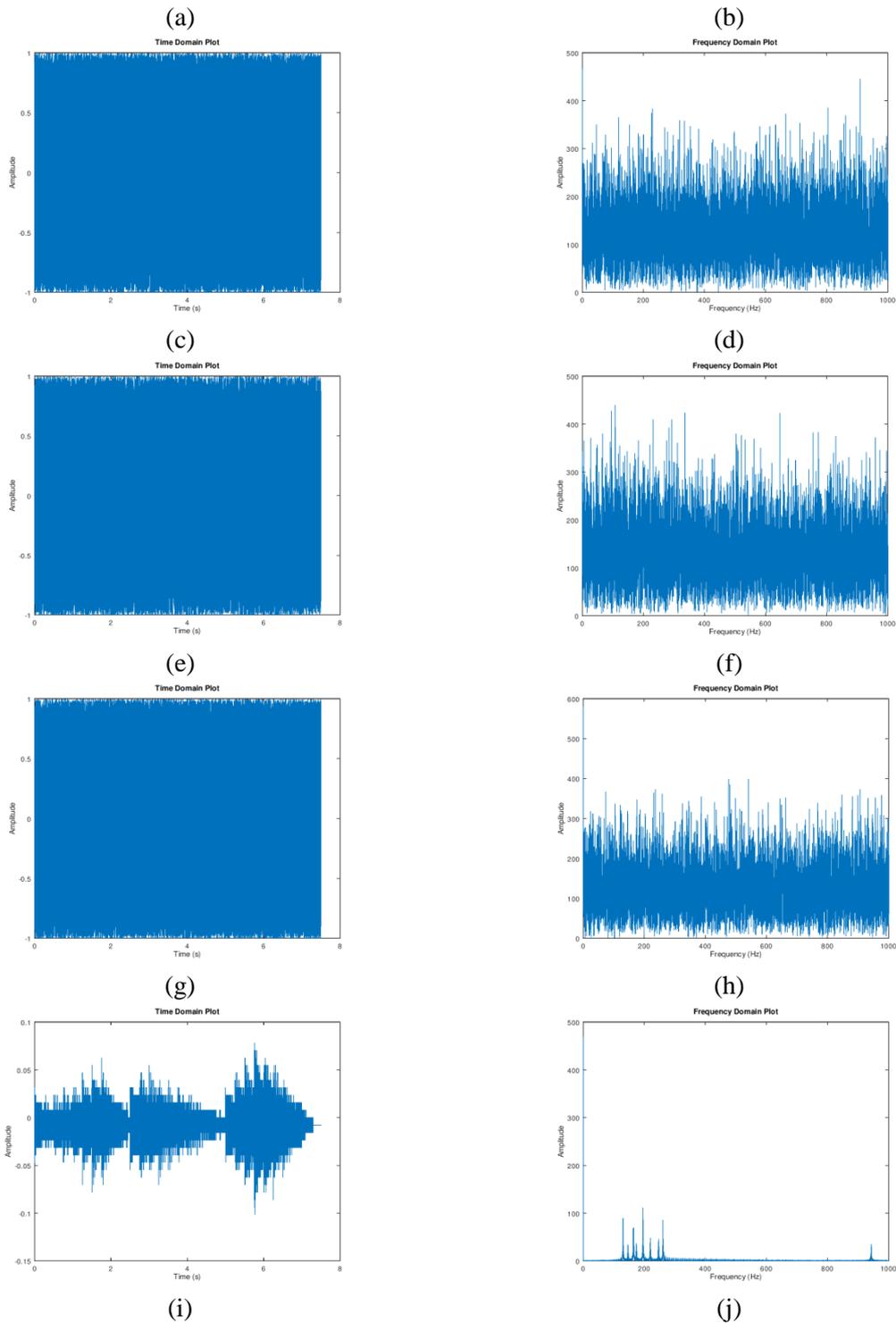

**Fig. 6** Demonstration of our proposed cycle attack: (a,c,e,g,i) time-domain plots of the original plaintext, its first encryption cipher, its second one, its third one and its fourth one respectively (b,d,f,h,j) their corresponding frequency-domain plots respectively

## 6. Lessons learned and recommendations for further works

In this section to follow, we outline some lessons learned from this cryptanalytic paper as well as some recommendations to be considered in improving the DECS-AU system or in designing another similar one. However, the DECS-AU system is considered at this stage to be faulty and with no security level, thus we recommend omitting it and don't use this system for any security concern. Some recommendations for designing similar systems are outlined below:

1) Make strong confusion/diffusion between key block and plain block chains, thus, it is recommended to use some cryptographic modes (e.g. CBC, CFB, OFB …) in order to diffuse plain blocks along with both cipher blocks and key blocks. Fig. 7 shows a diagram of CBC block chaining mode in which each ciphertext block depends on all previous plaintext blocks (IV is an initialization vector used to encrypt the first block) .
2) Use some existing DNA computing operations, we encourage the use of conventional DNA XOR operation as in [29,30].
3) Change the shuffling method and use more robust permutation maps. We propose for example to generate entirely key dependent permutation maps using other chaotic maps or any other methods to generate permutation maps. In addition to that, we recommend the use of both intra-block and inter-block permutations. That is, permuting bytes within a block and also from block to another one according to some permutation maps.
4) Use other chaotic maps since logistic map is not recommended for cryptographic use as shown in [32].

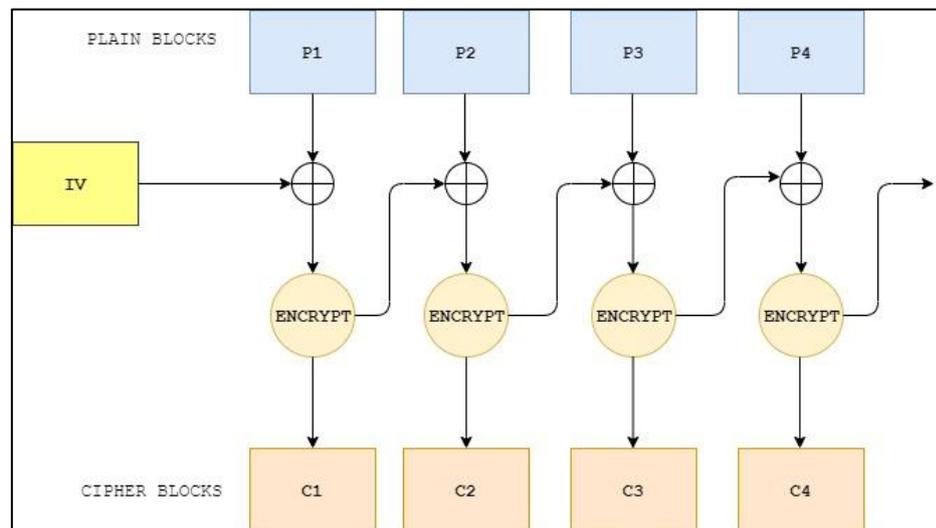

**Fig. 7** Diagram of CBC cryptographic mode

## 7. Discussion and Conclusion

In this paper, we analyzed the security of a new audio data cryptosystem proposed in [25] based on key chaining and DNA encoding. The key chaining algorithm aims to generate a new key block for every plain block. This process, as claimed by authors of the scheme in [25], allows enhancing the security of the system. Nevertheless, we found that this process has two drawbacks: the main and first one is depicted in (Section 3, vulnerability (1)) as this process don't take in consideration neither the plain block nor the cipher block to generates new key blocks, which leads to a bad confusion/diffusion implementation. The second one is that repeating the overall process of the key generation for every single plain block is time consuming. Some other vulnerabilities found in the scheme are detailed in (Section 3) including the period-two of the shuffling map which contributes in major part, to the deficiency of the system.

In addition to that, we have mounted three particular attacks (i.e. differential chosen plaintext attack, chosen ciphertext attack and cycle attack) to evaluate the system's security and its robustness. Negatively, we have found that the system is not as secure as claimed in [25]. The chosen ciphertext attack is the simplest one since it has lowest time complexity but it requires temporary access to the decryption machinery under the targeted key. The two other attacks are both considered as a chosen plaintext attack but they differ in some aspects. The first attack named differential chosen plaintext attack is actually based on two combined attacks: differential attack and chosen plaintext attack. The minimum number of required chosen plaintexts for this proposed attack is two chosen plaintexts. The second one called cycle attack is a type of chosen plaintext attack but it do not require any intervention from the part of the attacker such that programming new pieces of code, making some customized sets of plaintexts or do calculations… all the attacker needs is the temporary access to the encryption machinery and the ciphertext to be broken. That is, by re-encrypting any given ciphertext three successive encryptions we end up by recovering the original unknown plaintext. If needed, the key could be recovered (fully or partially) by applying a known-plaintext attack.

Our attacks' demonstrations are shown visually based on time-domain plot, frequency-domain plot or both of them. We have chosen to depict the most convenient plot type for every result to shed light upon the main differences between results and to be as representative as possible, for example for *sine* waves it is better worth to use frequency-domain plots to visualize the information data.

Our results show the weakness of the system under study and that it is not recommended for cryptographic use and information security. In addition to that, we have detailed our observed vulnerabilities and we have addressed some recommendations deduced from this cryptanalytic paper to avoid such vulnerabilities in further works.

**References**:

[1] Habib, Ibrahim W., and Tarek N. Saadawi. "Multimedia traffic characteristics in broadband networks." *IEEE Communications magazine* 30.7 (1992): 48-54.

[2] Du, Miao, et al. "Big data privacy preserving in multi-access edge computing for heterogeneous internet of things." *IEEE Communications Magazine* 56.8 (2018): 62-67.

[3] Li, Huining, et al. "A selective privacy-preserving approach for multimedia data." *IEEE multimedia* 24.4 (2017): 14-25.

[4] Ghadi, Musab, Lamri Laouamer, and Tarek Moulahi. "Securing data exchange in wireless multimedia sensor networks: perspectives and challenges." *Multimedia Tools and Applications* 75.6 (2016): 3425-3451.

[5] Janakiraman, Siva, et al. "Lightweight chaotic image encryption algorithm for real-time embedded system: Implementation and analysis on 32-bit microcontroller." *Microprocessors and Microsystems* 56 (2018): 1-12.

[6] Bensikaddour, El-Habib, Youcef Bentoutou, and Nasreddine Taleb. "Embedded implementation of multispectral satellite image encryption using a chaos-based block cipher." *Journal of King Saud University-Computer and Information Sciences* (2018).

[7] Lin, Zhihao, et al. "A novel fast image encryption algorithm for embedded systems." *Multimedia Tools and Applications* (2019): 1-21.

[8] Chen, Shuai, XianXin Zhong, and ZhengZhong Wu. "Chaos block cipher for wireless sensor network." *Science in China Series F: Information Sciences* 51.8 (2008): 1055.

[9] Gencoglu, Muharrem Tuncay. "Embedded image coding using laplace transform for Turkish letters." *Multimedia Tools and Applications* (2019): 1-14.

[10] Mishra, Dheerendra, et al. "Efficient authentication protocol for secure multimedia communications in IoT-enabled wireless sensor networks." *Multimedia Tools and Applications* 77.14 (2018): 18295-18325.

[11] Hraoui, S., et al. "Benchmarking AES and chaos based logistic map for image encryption." *2013 ACS International Conference on Computer Systems and Applications (AICCSA)*. IEEE, 2013.

[12] Bhargava, Bharat, Changgui Shi, and Sheng-Yih Wang. "MPEG video encryption algorithms." *Multimedia Tools and Applications* 24.1 (2004): 57-79.

[13] El-Latif, Ahmed A. Abd, Li Li, and Xiamu Niu. "A new image encryption scheme based on cyclic elliptic curve and chaotic system." *Multimedia tools and applications* 70.3 (2014): 1559-1584.

[14] Shakir, H. R. (2019). An image encryption method based on selective AES coding of wavelet transform and chaotic pixel shuffling. *Multimedia Tools and Applications*, 1-15.

[15] ur Rehman, A., & Liao, X. (2019). A novel robust dual diffusion/confusion encryption technique for color image based on Chaos, DNA and SHA-2. *Multimedia Tools and Applications*, *78*(2), 2105-2133.

[16] Claude Shannon, « Communication Theory of Secrecy Systems », *Bell System Technical Journal*, vol. 28, 4 octobre 1949, p. 662

[17] Alvarez, Gonzalo, and Shujun Li. "Some basic cryptographic requirements for chaos-based cryptosystems." *International journal of bifurcation and chaos* 16.08 (2006): 2129-2151.

[18] Li, Shujun, et al. "Cryptanalysis of the RCES/RSES image encryption scheme." *Journal of Systems and Software* 81.7 (2008): 1130-1143.

[19] Li, Chengqing, et al. "Cryptanalyzing image encryption using chaotic logistic map." *Nonlinear Dynamics* 78.2 (2014): 1545-1551.

[20] Li, C., Zhang, Y., & Xie, E. Y. (2019). When an attacker meets a cipher-image in 2018: A year in review. *Journal of Information Security and Applications*, *48*, 102361.

[21] Özkaynak, Fatih. "Brief review on application of nonlinear dynamics in image encryption." *Nonlinear Dynamics* 92.2 (2018): 305-313.

[22] Adleman, Leonard M. "Molecular computation of solutions to combinatorial problems." *Science* (1994): 1021-1024.

[23] Reif, John H. "Paradigms for biomolecular computation." *First International Conference on Unconventional Models of Computation, Auckland, New Zealand*. 1998.

[24] Zhang, Yunpeng, et al. "A DNA-Based Encryption Method Based on Two Biological Axioms of DNA Chip and Polymerase Chain Reaction (PCR) Amplification Techniques." *Chemistry–A European Journal* 23.54 (2017): 13387-13403.

[25] Naskar, Prabir Kumar, et al. "DNA Encoding and Channel Shuffling for Secured Encryption of Audio Data." *Multimedia Tools and Applications* (2019): 1-24.

[26] Iqbal, Sajid, et al. "Study of nonlinear dynamics using logistic map." *LUMS 2nd International Conference on Mathematics and its Applications in Information Technology (LICM08)*. 2008.

[27] Petitcolas, Fabien. "La cryptographie militaire." (1883).

[28] Genot, Anthony J., Jonathan Bath, and Andrew J. Turberfield. "Reversible logic circuits made of DNA." *Journal of the American Chemical Society* 133.50 (2011): 20080-20083.

[29] Yan, Hao, et al. "Parallel Molecular Computations of Pairwise Exclusive-Or (XOR) Using DNA "String Tile" Self-Assembly." *Journal of the American Chemical Society* 125.47 (2003): 14246-14247.

[30] Gupta, Vineet, Srinivasan Parthasarathy, and Mohammed Zaki. "Arithmetic and logic operations with DNA." (1997).

[31] Barua, Rana, and Janardan Misra. "Binary arithmetic for DNA computers." *International Workshop on DNA-Based Computers*. Springer, Berlin, Heidelberg, 2002.

[32] Arroyo, David, Gonzalo Alvarez, and Veronica Fernandez. "On the inadequacy of the logistic map for cryptographic applications." *arXiv preprint arXiv:0805.4355* (2008).